\documentclass[12pt]{article}   	
\usepackage{geometry}                		
\usepackage{graphicx}					
\usepackage{amssymb}
\usepackage{color}
\usepackage{authblk}
\usepackage{draftwatermark}
\SetWatermarkScale{4}

\title{An Algorithmic Approach to Forecasting Rare Violent Events: 
An Illustration Based in IPV Perpetration\thanks{Thoughtful comments 
and suggestions were provided by colleagues Aaron Chalfin, 
John MacDonald, and Greg Ridgeway.}}
\author[1,2] {Richard A. Berk}
\author[3] {Susan B. Sorenson}
\affil[1] {Department of Criminology, University of Pennsylvania}
\affil[2] {Department of Statistics, University of Pennsylvania}
\affil[3] {School of Social Policy and Practice, University of Pennsylvania} 

\begin{document}
\maketitle

\begin{abstract}
Mass violence, almost no matter how defined, is (thankfully) rare. Rare events are very difficult to study in a systematic manner. Standard statistical procedures can fail badly and usefully accurate forecasts of rare events often are little more than an aspiration. We offer an unconventional approach for the statistical analysis of rare events illustrated by an extensive case study. We report research whose goal is to learn about the attributes of very high risk IPV perpetrators and the circumstances associated with their IPV incidents reported to the police. Very high risk is defined as having a high probability of committing a repeat IPV assault in which the victim is injured. Such individuals represent a very small fraction of all IPV perpetrators; these acts of violence are relatively rare. To learn about them nevertheless, we apply in a novel fashion three algorithms sequentially to data collected from a large metropolitan police department: stochastic gradient boosting, a genetic algorithm inspired by natural selection, and agglomerative clustering. We try to characterize not just perpetrators who on balance are predicted to re-offend, but who are very likely to re-offend in a manner that leads to victim injuries. With this strategy, we learn a lot. We also provide a new way to estimate the importance of risk predictors. There are lessons for the study of other rare forms of violence especially when instructive forecasts are sought. In the absence of sufficiently accurate forecasts, scarce prevention resources cannot be allocated where they are most needed. 
 \end{abstract}

\section{Introduction}
Forecasts of risk are routinely made in a wide variety of situations. What is the probability that a hurricane will strike the Gulf Coast in a particular hurricane season? What is the probability that a given high school student will be accepted by his or her college of choice? What is the probability that a particular business firm will declare bankruptcy? Coupled with each probability is the expected cost should the event of concern occur. For the bankruptcy example, repayment of debt at 10 cents on the dollar means a loss of 90 cents for every dollar invested. Risk formally is defined as the costs of a particular event multiplied by the probability that the event will occur. 

Forecasts of risk can be useful if they lead to actions that are better informed. For undesirable outcomes, one hopes that prevention strategies can be implemented or that plans for remedial action after the fact can be made. This has long been well understood by criminal justice decision makers in the United States. Indeed, risk assessments have been used to inform criminal justice decisions since the 1920s (Burgess, 1928). One might wonder, therefore, whether forecasts of risk might be instructive for contemporary incidents of mass violence. Without good forecasts, scarce prevention and remedial resources easily can be misallocated. 

For almost any reasonable definition of mass violence, constructing sufficiently accurate forecasts is a daunting undertaking. This holds whether one is trying to forecast the likely perpetrator, location, or timing of an event. One obstacle is that mass violence is very heterogeneous. It can include, school shootings, homicides committed by disgruntled employees, brutal hate crimes, systematic execution of witnesses at a crime scene, fatal assaults by perpetrators of intimate partner violence (IPV), and other mass violence in which the motives are obscure (e.g., the October, 2017 Las Vegas music festival mass shooting in which 58 people were killed and 851 were injured). Although understanding mass violence in general is an admirable aspiration, in the medium term at least, different forms of mass violence might be productively examined separately. Useful forecasts will probably require different approaches for different kinds of mass violence because the risk factors and their importance will likely vary. 

Another obstacle is achieving a consensus about what the most relevant observational units should be. One important distinction is between the settings in which the violence occurs and the people found in these settings. Does one want a  forecast for a school as a whole or a forecast for each student in that school? Likewise, should the observational units be businesses or their individual employees? What about places of worship versus individual members of their congregations? In addition, the setting may be a kind of event rather than a place. For example, the observational units may be armed robbery incidents or rock concerts. 

Even if clear definitions for different kinds of mass violence could be provided and, for each, sensible observational units specified, a third obstacle is very low base rates. One consequence is that the raw numbers of such events will be small, often in no more than double digits. For example, one very large metropolitan area along the I-95 corridor had in a recent year fewer than 10 homicides related to intimate partner violence, and for none were there more than 3 victims; most had a single victim. One would need to accumulate intimate partner homicides from across the country to arrive at a mass violence total of more than a half dozen incidents. There is not much information that can be extracted from so few observations, especially when one might hope to learn what risk factors distinguish IPV mass violence incidents from the hundreds of thousands of IPV incidents in which no one dies. 

A more subtle concern is that very low base rates lead to very accurate but trivial forecasts. For example, Time magazine reports that in 2018, there were a total of 17 ``school shootings'' in the United States (Wilson, 2018). Suppose during that year there were about 100,000 public schools in the United States (National Center for Education Statistics, 2018). The probability that any given public school will victimized by a school shooting in that year is about .00017. If one had forecasted that for any given school there will be no shootings in 2018, that forecast would have been correct with a probability of over .999 using no risk factors whatsoever. It is hard to imagine that any forecasting procedure using risk factors could do better. If one proceeded nevertheless with standard statistical tools, it is likely that no useful risk factors would be identified. The numerical methods used would rapidly conclude that nothing to improve forecasting accuracy could be found. So why bother?\footnote
{
It might seem that statistical modeling using extreme value distributions could solve the problem (Coles, 2001). From a given type of extreme value distribution, one has the ability to extrapolate to rare events in the tails. But even before getting into a number of difficult details, one must know at least the form of the extreme value distribution, and in order to forecast, how to include the role of predictors. In other words, one has a very demanding model specification problem for phenomena that currently are poorly understood. One important risk is that untestable assumptions will be introduced to justify a particular specification; what some disparagingly call ``assume and proceed statistics.''
}

The answer is lies in the costs of mass violence. Although mass violence is rare, it can have devastating consequences. In addition to the tragic loss of life and the grieving of family members and friends, mass violence can undermine trust in government institutions to guarantee public safety. Mass violence also can weaken confidence in appointed and elected public officials and elicit racial, ethnic and religious scapegoating. For these reasons and others, efforts to reduce mass violence can be terribly important. Risk forecasting can help, at least in principle. In this paper consider ways to estimate the probabilities of mass violence assuming that the costs will be large by almost any metric.

Hence, the challenge. Effective forecasts of mass violence may prove to be useful, but the statistical obstacles are formidable. Herein, we will illustrate the potential of a novel approach to forecasting rare events. In part because of our access to unique data, the test bed is incidents of intimate partner violence in which the victim sustains injuries. Such incidents are usually not crimes of mass violence, but, as a form of intentional violence, raise many of the same statistical difficulties. In particular, for a typical set of IPV incidents, cases in which the victim is injured are rare. 

\section{IPV Risk Assessment With Low Base Rates}
Like most criminal justice risk assessments, risk assessments for intimate partner violence (IPV) typically use very broad definitions of the forecasting target. Often the forecasting target is simply the presence or absence of any actions that qualify under existing statutes. A loud argument can suffice. At the other extreme can be a lethal assault. Consequently, the usual search for risk factors can be compromised by very heterogeneous outcomes. An important risk factor for an argument may be an unimportant risk factor for an assault resulting in injuries. 

A few studies have narrowed their focus to very serious forms of intimate partner violence in which the victim is injured or even killed. Such outcomes make the research extremely important. But to be effective, the research must overcome very low base rates making identification of risk factors immensely difficult. 

In the pages ahead, we address and try to circumvent the problems caused by low base rates for IPV in which the victim is injured.\footnote
{
One might think that a good meta-analysis could provide a solution to the low base rate problem (Spencer and Stith, 2018). But at best, the only gain would be statistical power. As discussed below, a low base rate can undercut the estimated contributions of all predictors because it is very difficult to fit the data better than the marginal distribution of the highly unbalanced response. A forecast for the more common outcome will be correct most of the time with no help from the predictors whatsoever. There also are a variety of technical obstacles, especially when meta-analysis is applied to observational data (Berk, 2007).
} 
Using a unique dataset, we focus on the attributes of very high risk IPV perpetrators and the circumstances associated with their IPV incidents reported to the police. Very high risk is characterized as having a high probability of committing a repeat IPV assault in which the victim is injured. 

Rather than rely solely on a conventional data analysis of IPV incidents, we apply three algorithms sequentially to data from a large metropolitan police department: stochastic gradient boosting, a genetic algorithm inspired by natural selection, and agglomerative clustering. The first is used to define a fitness function, the second is used to construct a population of very high risk IPV offenders, and the third is used to help visualize the results. The constructed population does not have a problem with low base rates and instructive results are obtained.

\section{Past Research}
The very large literature on risk factors for intimate partner violence can be organized into three groupings. Some studies try to construct a causal account in which risk factors are treated as causes. Pathbreaking work by Straus and Gelles (1990) is an excellent example. Abransky and colleagues (2011) provide an international example. Very recent research by Weitzman (2018) continues in this tradition, focusing on how greater educational achievement for women can reduce victimization. For our purposes, such work is peripheral because intimate partner violence typically is very broadly defined. For example, intimate partner violence can be primarily comprised of threats or can include serious injuries requiring medical care. These are treated as different manifestations of the same underlying phenomenon.

A second tradition uses risk factors to characterize the ongoing dangers faced by victims of intimate partner violence. This approach can be traced back to work by Campbell (1995), and has led to several important follow-up studies (Campbell et al., 2003; 2007; 2009). Storey and Hart (2014) provide a recent example of the strengths and weakness of this approach. Although the attention to very serious intimate partner violence, often homicide, fits within our goals, the concern with explanation rather than prediction does not. A risk factor that may help explain why a homicide is more likely may have little forecasting power. Perhaps the major hurdle for such research, however, is the very low base rate. Lethal intimate partner violence, although certainly tragic, is very rare. 

A final approach centers on forecasting, typically to help inform criminal justice actions (Berk, 2018; Berk et al., 2005; 2016; Cunha et al., 2016). There is usually no causal account because risk factors are evaluated primarily by how much they improve forecasting accuracy. The research cited can include intimate partner violence in which there are injuries or even fatalities, but it too is challenged by very low base rates.  

\section{Data}
For all domestic violence dispatches confirmed as domestic violence cases by arriving officers, a special offense form was filled out. We had worked with the local police department to design the forms, which elicits a much wider range of information than what had previously been collected. (The form is still in use.) We were provided with a total of over 54,000 forms for the calendar year 2013. Each form characterized a domestic violence incident. 

Domestic violence was broadly defined, as is customary in law enforcement, to include disputes between parents and children, between siblings, and other variants on ``domestic,'' including intimate partners. We reorganized the data to include only incidents of intimate partner violence with the perpetrator as the analysis unit. Once irrelevant incidents were removed (e.g., a request for information only), there were 22,449 cases. For each perpetrator, we used the information in the earliest recorded incident in 2013 as our platform to forecast whether the victim in any \textit{subsequent} incidents in that year was recorded by police as having physical injuries. Approximately 20\% of the perpetrators in the initial incident had at least one subsequent IPV incident in 2013, and approximately 5\% had a subsequent IPV incident in 2013 in which the victim was injured. Repeat IPV incidents with reported victim injuries are quite rare, which presents a substantial data analysis challenge. Nevertheless, our response variable is the whether there is a subsequent IPV incident during 2013 in which the victim is injured. By ``subsequent'' we mean chronologically later than the IPV incident in 2013 that is the source of the baseline data. Further details about the data are provided in Small et al., 2019.

We selected all predictors from the collected offense forms for each perpetrator's initial IPV incident in 2013.  The far left column of Table~\ref{tab:predictors} shows the predictors used. All are indicator variables coded so that a ``1'' represents the presence of the attribute and a ``0'' represents the absence of the attribute. It will later help conceptually if one thinks of a ``1'' as switching a predictor on and a ``0'' as switching a predictor off. We will have much more to say about the predictors shortly.

\section{Methods}

As an initial benchmark and to motivate our statistical approach, a conventional logistic regression was applied to the data. Poor performance was expected. Because 95\% of the perpetrators did not commit a new, reported IPV incident in which the victim was injured, one can predict using no predictors that such an incident will not occur and automatically be right 95\% of the time. One cannot expect logistic regression to do any better.  

\begin{figure}[htbp] 
   \centering
   \includegraphics[width=4in]{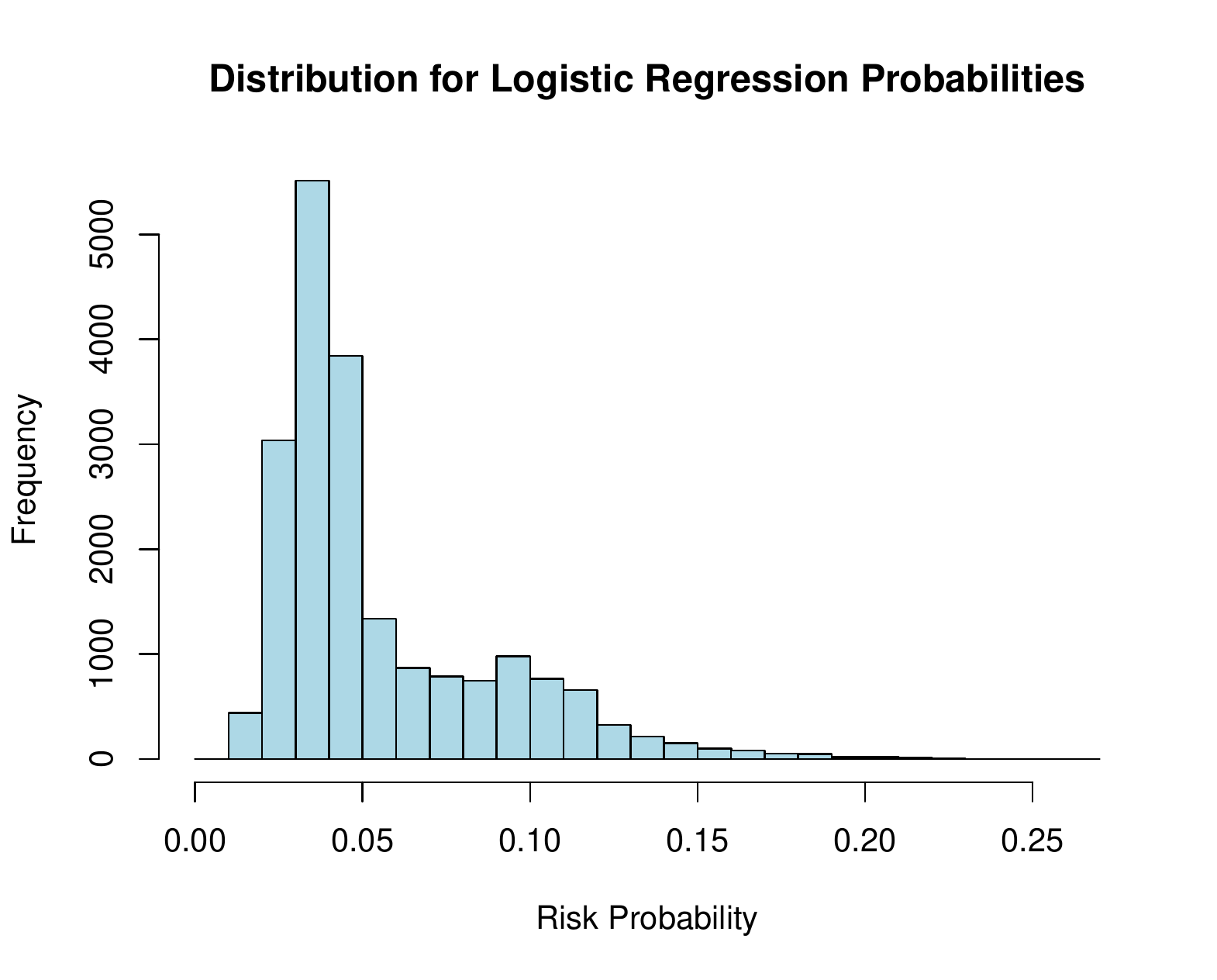} 
   \caption{Risk Probabilities from A Logistic Regression}
   \label{fig:logistic}
\end{figure}

Figure~\ref{fig:logistic} illustrates the problem. The mass of fitted probabilities fall around .04, and none exceed .27. The highest risk perpetrator had less than a 30\% chance of re-offending. With no fitted probabilities larger than .50, no perpetrators would be forecasted to be reported for a new IPV incident in which the victim was injured.

Even with such difficult data, machine learning procedures can do better (Berk and Bleich, 2013). We applied stochastic gradient boosting (\textit{gbm} in R).\footnote
{
Greg Ridgeway is the initial and primary author of \textit{gbm}.
}
Building on past forecasting studies of domestic violence that used machine learning (Berk et al., 2016), our target cost ratio treated false negatives as 10 times more costly than false positives. In other words, it was 10 times worse failing to correctly classify an IPV incident with injuries than failing to correctly classify an IPV incident with no injuries. 

We use the cost ratio as a place holder. When working with stakeholders on real applications, the cost ratio becomes a policy preference they would need to specify. But the 10-to-1 ratio is plausible. In any case, for the analyses that follow, the target cost ratio is peripheral to our main concerns. 

The data were randomly divided into training data having 20,000 observations and test data having 2449 observations. For reasons that will be apparent later, we will lean far more heavily on the training than the test data, which justifies the substantially larger number of training observations. There is no other formal rationale for our setting (cf. Faraway, 2016).

Because the outcome was binary, we used the conventional Bernoulli distribution to define the boosting residuals. We retained all of the \textit{gbm} default settings for the tuning parameters except that interaction depth was set at 10 to help capture the rare outcome events we were seeking. Reasonable variation in the tuning parameters (e.g., an interaction depth of 6) made little difference. The number of iterations was determined by 5-fold cross validation. There are difficult technical problems with cross-validation, but it seems to perform well in practice (Hastie and Tibshirani, 2009, Section 7.10.). Such performance is very important for boosting, which can badly overfit outcome probabilities (Mease et al., 2007).

The primary boosting output of interest was the learned fitting function that can be used to construct fitted values, conventionally treated as probabilities. Here, they convey the risk of a perpetrator committing a new IPV incident in which the victim is injured. As discussed shortly, the fitted probabilities ranged from a little more .3 to a little less more than .70. About a quarter of the perpetrators were predicted to re-offend  in a manner leading to victim injuries (i.e., fitted probability $ > .50$). 

But these results also were unsatisfactory. Even with the 10-to-1 cost ratio, the results were dominated by the perpetrators who did not re-offend because it was so easy to fit those cases accurately. Moreover, many predictors that might have been useful for identifying the repeat perpetrators could not be discerned because the unbalanced response variable precluded them from having much impact on the boosting loss function. In addition, the predictor values for the repeat, violent offenders, were limited to those in the data set. There could well be many other kinds of very high risk perpetrators with the same predictors but with different configurations of predictor values. 

For these reasons, we extended the analysis strategy. We applied a genetic algorithm (Luke, 2013: Chapter 3; Mitchell, 1998) with the goal of constructing predictor profiles that a variety of very high risk offenders \textit{might} have. We were not limited to the actual predictor values for such offenders who were in our data. Insofar as a substantial number of very high risk, hypothetical perpetrators ``survived,'' it would be possible to determine which predictors and predictor values were responsible. Put in other terms, using the genetic algorithm, we sought to construct a new population of very high risk, violence prone perpetrators that could be studied in the same way one would study an observed, empirical population. In this manner, we hoped to circumvent, at least in part, the statistical problems caused by low base rates. Genetic algorithms have been applied in economics, population genetics, ecology, immunology, and biology (Mitchell, 1998: section 1.8);  our application is novel. 

We applied the \textit{GA} procedure in R.\footnote
{
GA was written by Luca Scrucca.
}
The earlier \textit{gbm} prediction module developed from training data served as the fitness function. Perpetrators with larger predicted probabilities to violently re-offend were defined as more fit. This meant that if predictors for members of the hypothetical population of very high risk offenders were used as input data for the learned boosting results, the predicted probabilities of repeat violence would be well over .50. 

100 populations, each with 500 perpetrators, were constructed sequentially by the genetic algorithm, although there was little improvement after about the 20th population. Just as with \textit{gbm}, we found that the default values for the tuning parameters worked quite well. There was no meaningful improvement when the tuning parameters were varied within reasonable values. In the end, we had a single Frankensteinian population of 500 unusually unsavory perpetrators.

Finally, we sought to characterize the most important predictors of repeat IPV when the victim is injured and how those predictors were related to one another. In addition to some simple calculations for the population of 500, we applied agglomerative clustering algorithm (Kaufman and Rousseeuw, 2005, Chapter 5), a form of unsupervised learning (\textit{agnes} in the \textit{cluster} library in R). The clusters produced confirmed our earlier conclusions in an easily understood visualization. We also devised a new way to estimate predictor importance.
 
\section{Boosting Results}

Table~\ref{tab:confusion} shows a conventional machine learning classification table for the test data constructed from the output of the stochastic gradient boosting application.\footnote
{
A machine learning classification table is often called a ``confusion table." It is a cross-tabulation of the true categorical outcome by the fitted categorical outcome. 
}
Even with an achieved 10-to-1 target cost ratio (i.e., $763/77=9.9$), correctly classifying the rare cases was difficult. The row labeled ``Actual Injuries'' shows that 47\% of the cases in which there was a repeat IPV incident with injuries were incorrectly classified. Although a dramatic improvement over the logistic regression results, the classification error rate for the rare events is hardly inspiring. The misclassification rate when there are no injuries is smaller (33\%), but because injury-free IPV is by far the most common outcome, the classification task is much easier. 

Turning from classification to forecasting shown in the columns of Table~\ref{tab:confusion}, the target 10-to-1 cost ratio led to over 750 false positives that, in turn, resulted in a forecasting error for repeat violence of 92\%; forecasts of repeat  violence would be wrong 92\% of the time.\footnote
{
With the 10-to-1 cost ratio, false positives were, as a policy matter, very cheap. It is then no surprise that the boosting algorithm works very hard to avoid false negatives, but not false positives. Indeed, it is happy to trade a substantial number of false positives for fewer false negatives. Should this tradeoff be unacceptable to stakeholders, the cost ratio is easily changed.
} 
We do far better with forecasts of the absence repeat violence, but an error rate of 4\% is only slightly better than the 5\% error rate one would obtain by simply applying a Bayes classifier to the marginal distribution of the response variable. In short, compared to the marginal distribution of the the response variable, the predictors don't help much if the goal is more accurate forecasts, although the improvement for incidents in which there were no injuries could be somewhat greater if the costs ratio were increased.\footnote
{
The preferred cost ratio is implemented as a special form of weighting. Therefore, one might think that some form of weighting could be used to solve the low base rate problem; just give the rare events more weight in the analysis. Even if such weighting could be justified by subject matter considerations, any apparent gains could be misleading. No new information is being added. Suppose there are 50 rare events, and one gives them double the weight. This is the same as counting each rare event twice. Each rare event has an exact duplicate and as just illustrated, one is asking for a substantial increase is false positives. In Table~\ref{tab:confusion}, use of asymmetric costs was introduced to distribute the false negative and false positive classification errors in a way that was consistent with specific policy preferences. That is a different problem. 
} 

\begin{table}[htp]
\caption{Stochastic Gradient Boosting Classification Table Using Test Data}
\begin{center}
\begin{tabular}{|l|c|c|c|} \hline \hline
  & Forecast No Injuries & Forecast Injuries & Classification error \\
  \hline
  Actual No Injuries & 1542 & 763 & 0.33 \\
  Actual Injuries & 77 & 67 & 0.47 \\
  \hline
  Forecasting Error & 0.04 & 0.92 &  \\
  \hline \hline
\end{tabular}
\end{center}
\label{tab:confusion}
\end{table}

\begin{figure}[htbp] 
   \centering
   \includegraphics[width=4in]{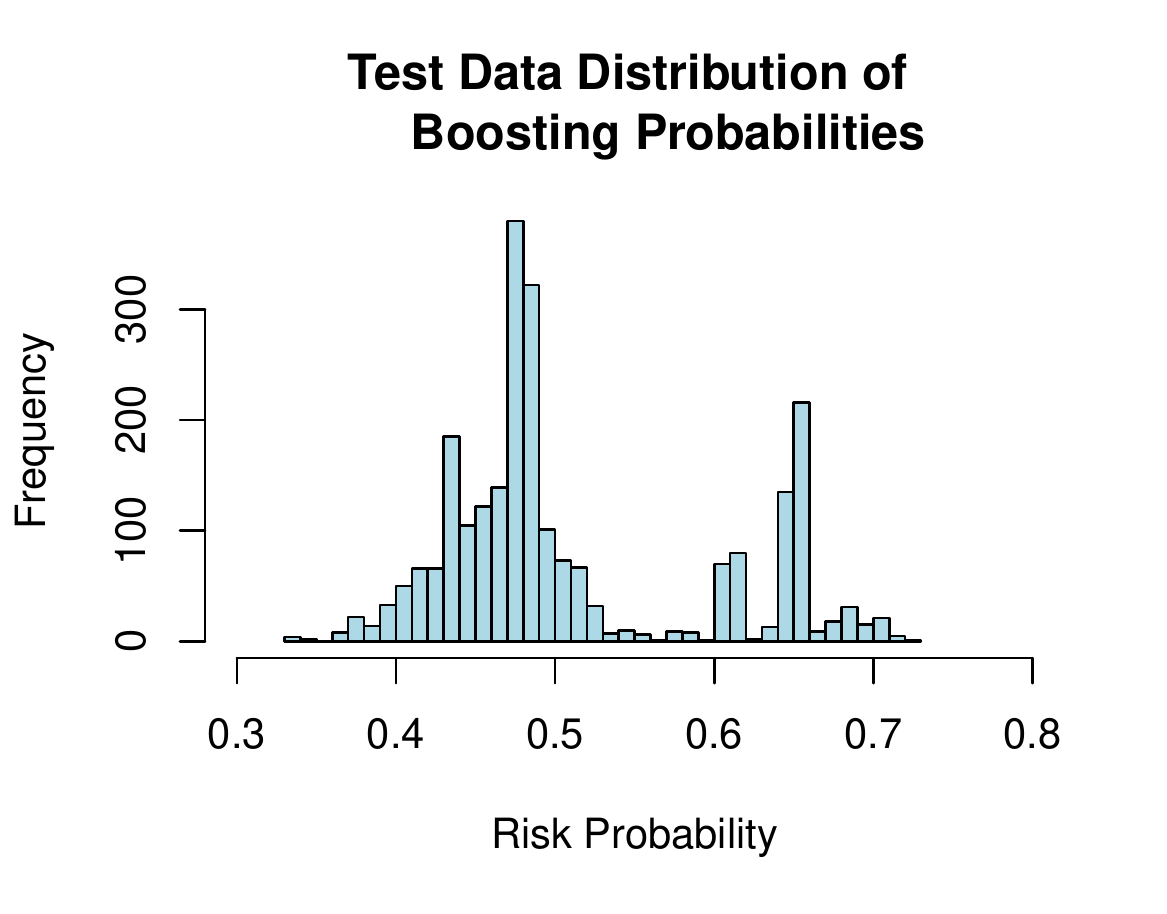} 
   \caption{Risk Probabilities from Test Data for Stochastic Gradient Boosting}
   \label{fig:boostrisk}
\end{figure}

For our purposes, far more instructive are the fitted values. Figure~\ref{fig:boostrisk} shows that the fitted probabilities range from a little more than .30 to a little more than .70. Clearly, there is a dramatic improvement over the fitted values from the logistic regression. A substantial number of fitted probabilities are to the right of .50, leading to forecasts of victim injuries. However, from Table~\ref{tab:confusion}, we know that the majority of these forecasts are false positives. Moreover, probabilities larger than .50 vary widely with a small minority having the among the highest risk probabilities. Finally, because the probabilities range widely, perhaps there is a range of etiologies leading to IPV injuries. Perpetrators forecasted to injure their intimate partners can differ substantially from one another. For example, some may on occasion lose control, perhaps under the influence of drugs or alcohol. Some may use violence on a routine basis systematically to enforce domination.

\begin{table}[htp]
\caption{Predictor Importance Forecasting Victim Injuries}
\begin{center}
\begin{tabular}{|l|c|c|c|}
\hline \hline
Candidate Predictors & In-Sample & Commonality  & Switched On or Off\\
                                 & Importance & Importance &  or In Between\\
\hline
Follow up $>$ 3 months & 27.12 & 1.00 & {\color{red} { {\color{red} {Always On}}}} \\
Prior DV Reports & 4.77 &1.00 &  {\color{red} {Always On}} \\
Victim $<$ 30 & 3.64 & 1.00 &  {\color{red} {Always On}} \\
Contact Information Given to Victim & 3.48& 0.00 &  {\color{blue} {Always Off}} \\
Offender Arrested & 3.23 & 1.00 &  {\color{red} {Always On}}\\
Offender $<$ 30 &  3.18& 1.00 &  {\color{red} {Always On}}\\
Victim Frightened & 3.13 & 0.43 &   {\color{green} {In between}} \\
Offender Polite & 3.08 & 0.00 &   {\color{blue} {Always Off}}\\
Currently Married  & 2.70 & 0.73&  {\color{green} {In between}} \\ 
Offender Cooperative & 2.67& 0.62 &  {\color{green} {In between}}\\
Offender White  & 2.66 & 0.20 &  {\color{green} {In between}}\\
Victim Shaking & 2.66 & 0.44 &  {\color{green} {In between}}\\
Visible Injuries & 2.61& 0.49 &  {\color{green} {In Between}} \\
Offender Black & 2.50& 1.00 &  {\color{red} {Always On}}\\
Victim Latina &  2.44 & 0.68 &  {\color{green} {In between}}\\
Victim Crying & 2.34 & 0.49 &  {\color{green} {In between}}\\
Children Present &  2.33 & 0.49 &  {\color{green} {In between}} \\
Offender Threatened & 2.22 & 0.26 &  {\color{green} {In between}}\\
Weapon Used & 2.17 & 0.26 &  {\color{green} {In Between}}\\
Former Relationship & 2.10 & 0.31 &  {\color{green} {In Between}}\\
PFA Ever & 2.02 & 1.00 &   {\color{red} {Always On}}\\
Offender Angry & 1.95 & 0.42 &   {\color{green} {In between}}\\
Offender Apologetic & 1.82 & 0.12 &  {\color{green} {In between}}\\
Furniture in Disarray &  1.71 & 0.00 &  {\color{blue} {Always Off}}\\
Relationship Breaking Up & 1.54 & 0.32 &  {\color{green} {In between}}\\
Victim's Clothes in Disarray & 1.47& 0.76 &  {\color{green} {In between}}\\
Evidence Collected &  1.38 & 0.76 &  {\color{green} {In between}}\\
Offender Stalked &  1.23 & 0.47 & {\color{green} {In between}} \\
Offender Strangled & 1.17 & 0.63  &  {\color{green} {In between}}\\
Taken to Hospital &  1.13 & 0.36 &  {\color{green} {In between}}\\
Formerly Married &  1.12 & 0.35 & {\color{green} {In Between}}\\
Statements Taken from Kids & 1.07& 0.22 &  {\color{green} {In between}} \\
Offender Broke In & 0.74 & 0.22 &  {\color{green} {In between}}\\
PFA Expired & 0.57 & 0.51 &  {\color{green} {In between}}\\
 \hline \hline
\end{tabular}
\end{center}
\label{tab:predictors}
\end{table}

The far left column of Table~\ref{tab:predictors} shows all of the predictors in order of their fitting importance according to \textit{gbm}. Fitting importance for a given predictor is defined as the average reduction in the deviance over the boosted regression trees used by \textit{gbm}. That is, importance is the in-sample, average contribution to the fit of the data. The second column shows the importance of each predictor as a proportion of the total deviance accounted for over all predictors. We call this ``in-sample importance.''\footnote
{
Ideally, predictor importance would be computed in the test data, which would provide a measure of ``out-of-sample importance'' (Berk, 2018).
}

By far, the most important predictor is whether the initial IPV incident occurred in the first 90 days of 2013. Then, the followup period was between 9 and 12 months. A partial dependence plot showed the relationship to be positive. The likely explanation is that perpetrators who entered the study early in the year had more time to re-offend. This is of little substantive interest, and serves as a sanity check on the boosting results. 

The second most important predictor is whether there had been prior domestic violence reported to the police. The relationship is also positive and also not a surprise. However, the relationship is weak. Even weaker is whether the offender is under 30 years of age. The relationship is positive. When the offender is under 30, the chances of a repeat incident with injuries are increased. 

One can certainly proceed farther in this fashion, but it is not clear what of practical use is being learned. Beyond the single most important predictor, the fraction of the fitted deviance attributed to each predictor is very small and often does not materially differ from one predictor to another. One does not even have a meaningful rank ordering.\footnote
{
The procedure is \textit{stochastic} gradient boosting and additional randomness is introduced by the random subsetting into training and test data. Re-running the analysis several times from the beginning produced classification tables almost the same as Table~\ref{tab:confusion}, but the importance measures shuffle the order of all but the single most important predictor.
}
The predictor importance measure by used by \textit{gbm} (i.e., the average standardized fraction of deviance ``explained'') also provides no insight into how forecasts or their accuracy are affected. There are better measures associated with other machine learning procedures (e.g., Brieman, 2001). Finally, boosting is not a model so that causal inferences are unjustified whatever the importance measure computed (Berk, 2018).

For this analysis, an additional problem is that the importance measures in Table~\ref{tab:predictors} are produced by a fitting exercise in which it is extremely difficult to reduce classification accuracy beyond applying a Bayes classifier to the marginal distribution. Many potentially important predictors for identifying high risk offenders may not surface. Moreover, imposing an outcome class depending on which side of .50 a risk probability falls obscures that there is range of values above .50 that could represent a variety different perpetrator types and true IPV risk. Finally, the measures of importance are not immediately responsive to one of our motivating questions: what do very high risk offenders have in common and how are such attributes related? 

\section{Genetic Algorithm Results}  

For those questions, we turn to the results from the genetic algorithm whose output must be understood into the context of genetic algorithmic machinery. For this application, the values of indicator variables are randomly altered with no regard for whether certain \textit{combinations} of such values make subject-matter sense. The algorithmic fitness function does not automatically weed out unlikely or even impossible combinations of predictor values because all that matters is the probability of violent re-offending. It has no inkling about predictor combinations of perpetrator values that are actually not possible in reality. The result can be  ``unicorns,'' interesting perhaps, but ultimately not real. For example, there could be IPV outcomes in which statements are taken from children in households where there are no children reported. Such potential problems are exacerbated by errors in the offense reports.

\begin{figure}[htbp] 
   \centering
   \includegraphics[width=4in]{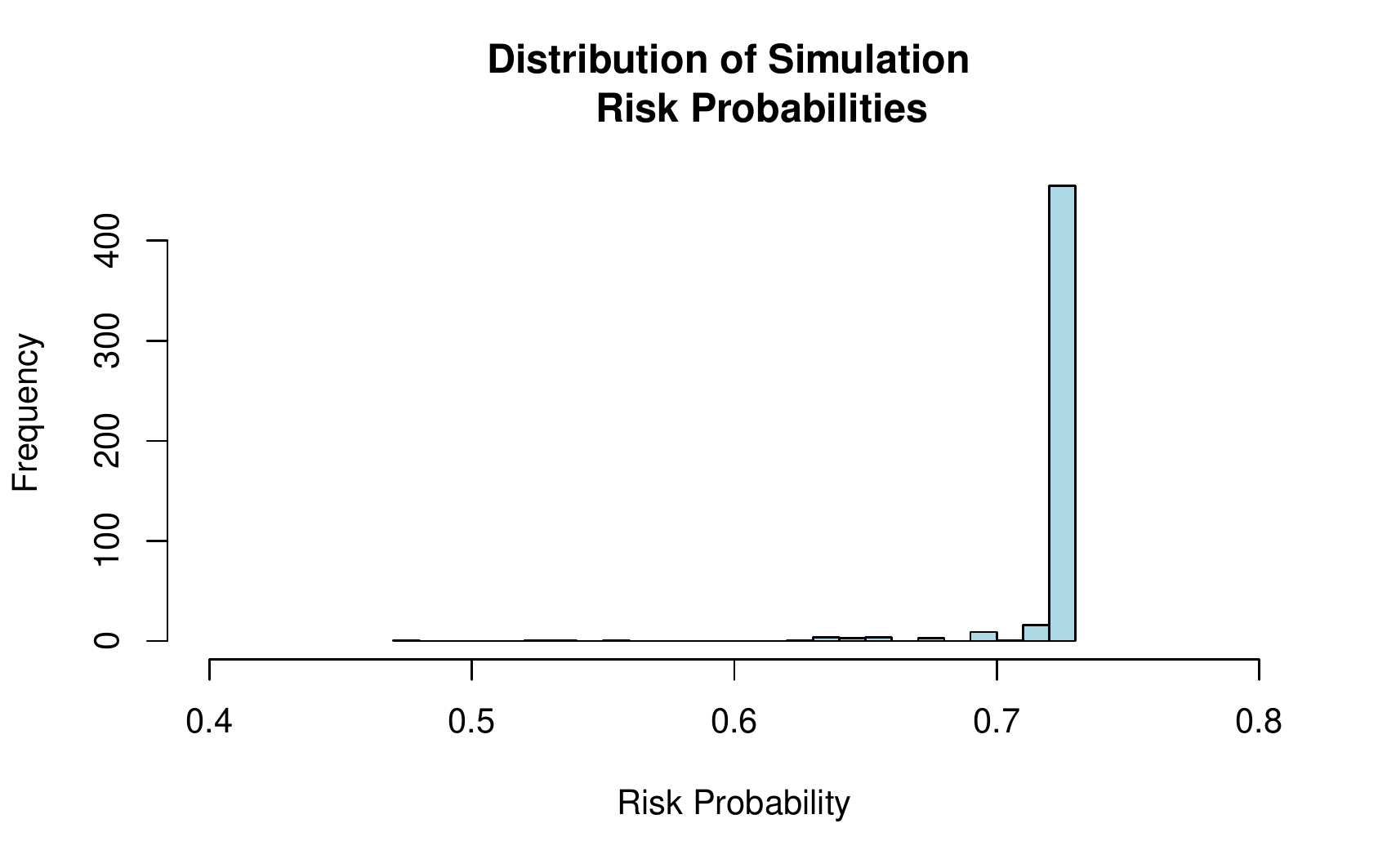} 
   \caption{Risk Probabilities from the Genetic Algorithm}
   \label{fig:garisk}
\end{figure}

Figure~\ref{fig:garisk} shows that we now have a hypothetical population of 500 \textit{composed almost entirely of very high risk perpetrators}.\footnote
{
The probability that an indicator predictor would have its value flipped as an offspring was produced (a ``mutation'') was set to .10. Changing it to .05 or .25 made no important difference except for altering the number of populations needed before no further improvement was found. The probability of a crossover (``sexual'' reproduction) between a random pair of perpetrators when an offspring was produced was set at .80. Dropping that value as low as .10 did not change the results in an important way, although again, the number of populations needed changed somewhat. The default crossover method was a ``single point'' procedure in which, for a single randomly chosen perpetrator, all predictor values for columns to the right of a randomly chosen column are swapped with the values for the same columns for another randomly chosen perpetrator (Umbarkar and Smith, 2015, Section 2.1).  By default, the fittest 5\% of the cases automatically survived to the next generation with no changes. Many of the background details can be found in Scrucca, 1993, published the \textit{Journal of Statistical Software}.
}
Almost all have a risk probability of .70 or larger. We now ask: what do these perpetrators have in common?

We computed for each predictor the proportion of the simulated population for which that predictor was switched on (i.e., that the predictor value was equal to 1). For example, should, for a given predictor, that proportion be 1.0, all members of the very high risk population have that predictor switched on. Should that proportion be, say, 0.35, 35\% members of the very high risk population have that predictor switched on. Should that proportion be 0.0, all members of the very high risk population have that predictor switched off. We call these proportions ``commonality importance.''

The third column in Table~\ref{tab:predictors} shows the commonality importance of each predictor. Values range widely. Some have a value of 1.0, which means that for this population they are always switched on. These are: 
\begin{enumerate}
\item
Follow-up $>$ 3 months;
\item
Prior DV Reports;
\item
Offender $<$ 30 Years Old;
\item
Victim $<$ 30 Years Old; 
\item
Offender Arrested;
\item
Offender Black; and 
\item
PFA Now or in the Past (i.e., a court-issued protection from abuse order)
\end{enumerate}

When switched on, each of these variables -- except arrest -- \textit{increase} the probability of a subsequent repeat incident with injuries reported to the police. The \textit{gbm}partial dependence plot shows that following an arrest, there is a slight \textit{reduction} of .04 in the probability of a repeat IPV incident with injuries. Yet, all of the perpetrators in the constructed, very high risk population had their arrest predictor given a value of 1.0; All were characterized by an arrest at the initial IPV incident. There seems to be a contradiction.  

Police in this jurisdiction are required to make an arrest for IPV incidents involving injuries (Pennsylvania Statute, Crimes and Offenses, chapter 27, section 2711, paragraph a). It follows that in reality, all of our high risk population would have been arrested at the initial, reported incident. Perhaps an arrest reduces their estimated risk, but not enough to remove them from the subset of very high risk subset offenders. We return to the role of arrest later; there is more to the story.\footnote
{
Recall that he partial dependence plots are constructed for all 20,000 perpetrators in the training data, not for the 500 very high risk offenders. 
}
We also will consider in more depth the role of race. Still, at this point, six of the seven predictors can be seen as potential aggravators.

Some predictors have a commonality importance value of 0.0, which means that for this population they are always switched off. These are:
\begin{enumerate}
\item
Victim Assistance Contract Information Given to the Victim (i.e., the contact information was \textit{not} given to the victim);\footnote
{
It is difficult to know what this predictor captures. Because all of the very high risk population would have been arrested, the offender would have been removed from the scene. Under these circumstances, the police might have understood that the department would actively follow up with the victim.
} 
\item
Offender Polite (i.e., the offender was \textit{not} police in his interaction with the police); and
\item
Furniture in Disarray (i.e., the furniture was \textit{not} in disarray.).\footnote
{
When furniture is in disarray, perhaps the victim's property rather than the victim herself is the perpetrator's target. For the initial incident, the victim is spared but the property destruction might convey a threatening message.
}
\end{enumerate}

From the partial dependence plots we learn that had any of these been switched on, the probability of a subsequent IPV incident in which the victim is injured is increased. Because they are all switched off for this population of perpetrators, they do not reduce subsequent violence. We think of the three predictors as potential mitigators. 

In short, Table~\ref{tab:predictors} identifies potential aggravators and mitigators for the risk of serious intimate partner violence. The archetypical perpetrator who injures his victim can be characterized by these aggravators and mitigators that often were not apparent in the boosting measures of importance. Other variables in the table can matter too, but they are not consistently switched on or off for this population of 500. 

Although the simulated population of very high risk offenders was generated from these data, it does not reproduce these data. There are no perfect matches. Across each of the 34 predictors, a maximum of 27 predicators had values that were the same for any perpetrator in the actual data and any perpetrator from the simulated population. That should be no surprise because very few perpetrators in the data had risk probabilities as high as those for the population of 500. 

One might wonder how many of these very high risk offenders would be false positives. There is no way to know because they are not included in the empirical data; there is, for the simulated population, no ``ground truth.'' But from those data, one can see that the number of false positives declines as the fitted risk probabilities increases beyond .50. For these very high risk offenders, most of the forecasted positives would probably be true positives.  

\section{Clustering Results}

\begin{figure}[htbp] 
   \centering
   \includegraphics[width=5in]{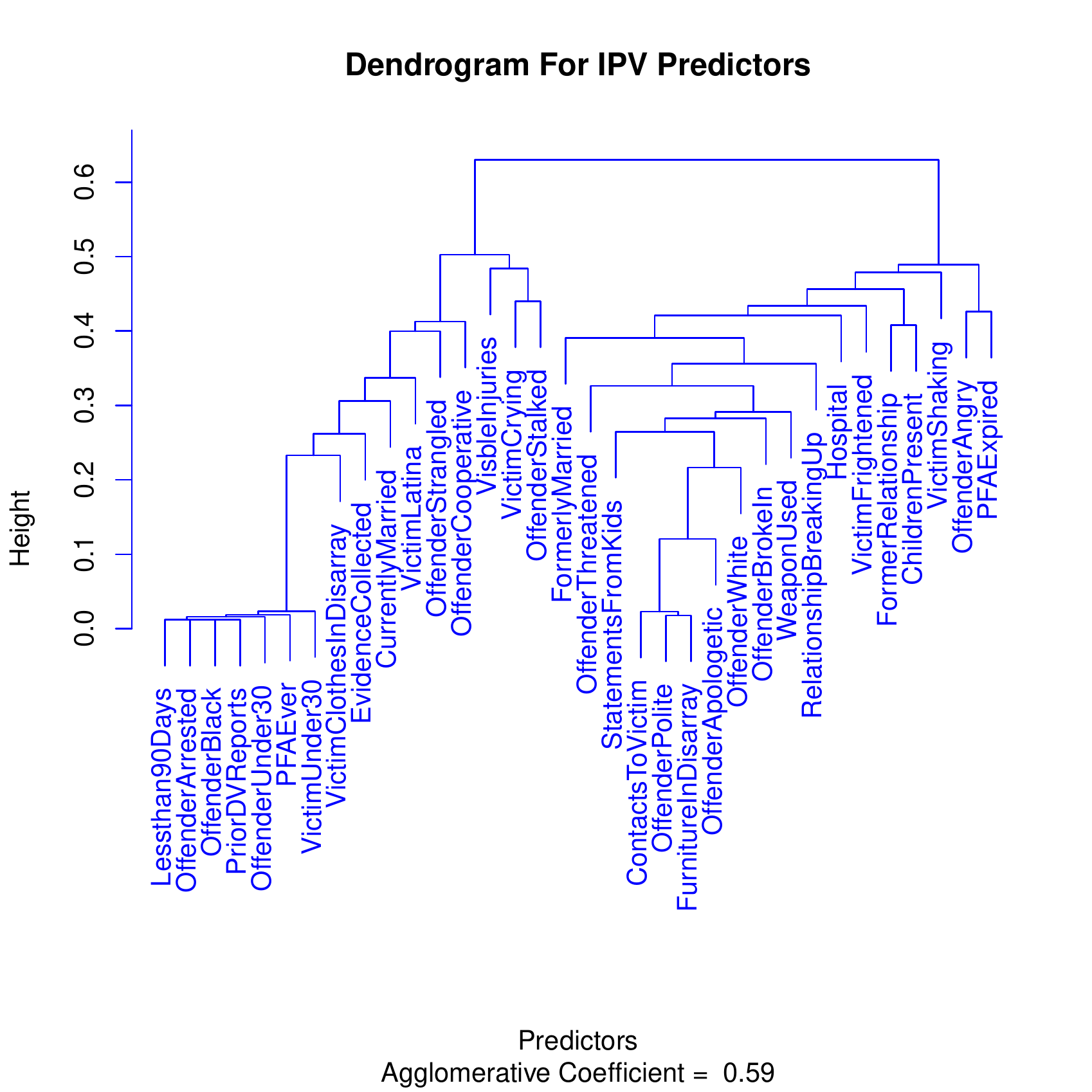} 
   \caption{Clustering of Injury IPV Predictors }
   \label{fig:cluster}
\end{figure}

In principle, a cluster analysis of the predictors should replicate the story extracted informally from Table~\ref{tab:predictors} and add instructive visualizations. Predictors with proportions of 1.0 should constitute one strong cluster, and predictors with proportions of 0.0 should constitute another strong cluster. Both extreme proportions imply maximum similarity, a bit like a correlation of 1.0 and -1.0.

Figure~\ref{fig:cluster} shows the dendrogram that results when an agglomerative clustering algorithm (Kaufman and Rousseeuw, 1990, Chapter 5)  is applied to the predictors in the very high risk population.\footnote
{ 
The process begins with the construction of dissimilarity matrix. Because the predictors are binary, the Gower method was used as a measure of similarity (Kaufman and Rousseeuw, 1990: 235-236). When transformed into dissimilarities, the index becomes a measure of distance between predictors. The algorithm starts with each predictor unclustered, finds the pair that is least dissimilar and combines them in a cluster. For each unclustered predictor in turn, a mean distance is computed between the within-cluster predictors and that predictor. In essence, the closest predictor is then taken into the cluster, and the process continues. Clusters are combined using similar reasoning based on the average distance between predictors in each possible pair of clusters.
}
``Height'' on the left margin of the figure is a standardized measure of dissimilarity within a cluster that ranges from 0 to 1 and necessarily gets larger when as one moves toward the top of the figure; the clustering begins at the bottom of the figure for the predictors that are the least dissimilar. The agglomerative coefficient at the bottom Figure~\ref{fig:cluster}, which also ranges from 0.0 to 1.0, is a standardized measure of cluster distinctness.\footnote
{
 The details are beyond the scope of this discussion. See Kaufman and Rousseeuw 1990: 211-212.
}
The moderate coefficient of .59 indicates that on the average, the clusters are quite distinct; the within-cluster dissimilarities are on the average substantially smaller than the between-cluster dissimilarities. 

The cluster on the lower left reproduces the ``always on'' predictors from the far right column in Table~\ref{tab:predictors}, but using the Gower measure not the importance commonality. Because the next predictor to join (Victim's Clothes in Disarray) is substantially higher (i.e., about .25), its distance from the cluster is meaningful. 

Finding that Black offenders increase risk merits examination. The baseline category is Hispanic offenders whose victims may be less inclined to call the police or cooperate once they arrive. This reinforces a point made earlier: the correspondence between IPV reported to the police and IPV not reported to the police can be far from perfect. We have no direct evidence that the findings for Black IPV perpetrators results from their victims being more inclined to report IPV incidents to the police. Moreover, Table~\ref{tab:predictors} documents that in the very high risk population we have some members who are both White and Black. As noted earlier, this could result from reporting errors on the offense forms, a coding errors when the forms were digitized, or an artifact of the genetic algorithm.\footnote
{
In principle, there are ways to avoid such anomalies. For example, cases with unrealistic configurations of predictor values could be dropped from the population. One could also alter the algorithmic code to prohibit some combinations of predictor values in the first place.
}
Whatever the cause, it undercuts any conclusions about the role of a perpetrator's race. 

The initial cluster to the right including Contacts To Victim, Polite Offender, and Furniture In Disarray always have a proportion value of 0.0. These predictors are always switched off. According to their partial dependence plots, each would decrease the chances of injury when switched on. Therefore, one would not expect them to be switched on for very high risk offenders. Because this cluster forms at about the same height as the cluster with predictors always switched on, both clusters have about the same, small within-cluster dissimilarity. If one wanted to enlarge the cluster, the next predictor to join would be whether the offender broke in. According to its partial dependence plot, it has almost no association with the IPV repeat incidents in which the offender is injured although it seems associated here with reductions in the probability injuries. Such events are so rare that it is difficult to even speculate on what the association might mean. 

Perhaps the most important conclusion from the clustering results is that selecting as important only those predictors that are universally switched on or off is discarding other useful information. The proportions in Table~\ref{tab:predictors} provide information that could also be exploited. It may be worth noting that the boosting importance measures and the commonality importance measures sometimes order predictors in the same way, especially for the predictors toward the very top of the table. But the predictors highlighted can differ substantially as well. 

\section{A New Approach to Predictor Importance}

A routine task in the development of criminal justice risk assessment tools is to document how each risk factor affects forecasts of risk. Typically, some risk factors will be alter risk forecasts more than others. Put another way, the credibility of risk forecasts depends substantially on the weight given each the predictor as risk scores are determined. We need to know how much each predictor ``moves the needle." None of our earlier results provide the information.

Working broadly from machine learning traditions formalized by Leo Breiman (2001), we developed new predictor importance measures implemented with the following steps. 

\begin{enumerate}
\item
Use the very high risk population of 500 as new data from which predicted probabilities are desired. Figure~\ref{fig:garisk} provides that information visually.
\item
Compute the mean risk probability to serve as a \textit{benchmark}. This number is the average risk probability when all of the predictors are set to the values determined by the genetic algorithm. 
\item
Construct new datasets, one for each of the predictors with a universal 1.0 or a universal 0.0. There will be 10 such datasets, each containing the full set of predictors (34 in our case). 
\item
For one of the 10 new datasets, select a predictor that is universally 1.0 or 0.0. Apply reverse coding for that predictor.\footnote
{
 If the predictor in question is universally a 1.0, recode it to 0.0. If the predictor in question is universally a 0.0, recode it to 1.0. There is no formal need to limit this process to predictors that are, for all members of the population, either equal 1.0 or 0.0. But these are the predictors that have the most promise of substantial importance because all 500 values will be recoded. If, for a given predictor, the proportion of 1s is, say, only .60, only 60\% of the values will be recoded. 
}  
For example, with the new dataset for the predictor Prior DV Reports, recode the 1s to 0s. \textit{All other predictor values are unchanged.} Repeat the reverse coding within each dataset, so that each predictor has its own dataset in which it alone has been reverse coded. 
\item
Using the boosting algorithmic structure found earlier, obtain the 500 predicted probabilities of a repeat IPV incident in which the victim is injured separately for each of the 10 datasets. There will be 10 sets of predicted probabilities.
\item
Compute the mean of the predicted risk probabilities separately for each dataset. 
\item
Compare each of these means to the mean probability when no predictors are recoded.\footnote
{
One could accomplish the same thing with single dataset (not 10) by reverse coding a given ``universal predictor,'' obtaining the desired estimate, reversing the reverse coding, and repeating the process for each universal predictor one at a time. It is here conceptually and operationally more direct to work with one dataset set for each universal predictor. 
}

\end{enumerate}

\begin{table}[htp]
\caption{Mean Risk Probabilities Computed by Reverse Coding}
\begin{center}
\begin{tabular}{|l|c|} \hline \hline
 Variable Recoded & Mean Risk Probability \\
 \hline
 None & 0.718 \\
Follow up $>$3 months: 1.0 to 0.0 & 0.431 \\
Prior DV Reports: 1.0 to 0.0 & 0.612 \\
Victim $<$ 30: 1.0 to 0.0 & 0.712 \\
Offender $<$ 30: 1.0 to 0.0 & 0.700 \\
Offender Arrested: 1.0 to 0.0 & 0.701 \\
Offender Black: 1.0 to 0.0 & 0.664 \\ 
Ever having a PFA: 1.0 to 0.0 & 0.709 \\
Contact information given to the Victim: 0.0 to 1.0 & 0.662 \\
Offender Polite: 0.0 to 1.0 & 0.683 \\
Furniture in Disarray: 0.0 to 1.0 & 0.642 \\
\hline \hline
\end{tabular}
\end{center}
\label{tab:impact}
\end{table}

Table~\ref{tab:impact} shows the results. The mean risk probability in the population of 500 is 0.718. This is the value of the benchmark. When each predictor is reverse coded (one at a time), the mean probability will drop in value. The larger the drop, the greater the impact that predictor has on the average risk.

Clearly, the largest impact by far is for the length of the follow up. The average risk probability drops from 0.718 to 0.431 when its values are recoded from 1.0 to 0.0. The next largest impact is far smaller. The average risk probability for Prior DV reports is 0.612 when its values are recoded from 1.0  to 0.0. Next, in order, the mean probability risk for furniture in disarray falls to 0.642 when its values are recoded from 0.0 to 0.1. If one proceeds through the table, the smallest impact is for victims under 30 years of age when its values are recoded from 1.0 to 0.0. The mean risk probability for victims under 30 years of aged is 0.712.\footnote
{
These measures of importance isolate the impact of the given predictor with all else held constant in the sense that none of the other predictors are recoded. This is in the spirit of an experiment; only the ``treatment'' is varied.
}

The reduction for the arrest variable is curious. From the boosting partial dependence plots, an arrest reduced the estimated probability of new IPV violence. But here we see that when an arrest is \textit{not} made, the probability of subsequent IPV resulting in injuries declines. If one or both of the results are not a product of some statistical artifact or errors in the offense forms, the impact of an arrest differs for the actual, rather heterogeneous population of perpetrators compared to the constructed, homogeneous, very high risk population. Alternatively, the arrest variable may be in part of a proxy for factors not measured in the dataset. In the very high risk population, for example, an arrest and a perpetrator's response to it, may deter a victim from reporting subsequent IPV incidents. 

Clearly, the universal predictors can reduce mean risk probabilities by meaningful amounts for this very high risk population. Less clear is whether \textit{differences} in impact between the predictors should be taken seriously. We re-ran the genetic algorithm several times and although by and large the same predictors were universally 1.0 or 0.0, the impact on predicted risk changed a bit, usually in the second or third decimal place. With the exception of the length of the follow-up period, this was sometimes enough to reorder the predictors in their importance.\footnote 
{
The very few times one of a universal predictor had less than all 1s or all 0s, the proportion of 1s or 0s was still very large. And on the rare occasion when a new universal predictor surfaced, it earlier had a large proportion of 1s or 0s. 
}
Modest re-rankings of this sort are not surprising because the genetics algorithm has random processes built it.  

The fitness function used by the genetic algorithm  is no doubt highly dependent on the training data and on the cost ratio imposed. Any thoughts of generalizing the findings for variable importance are surely premature. Yet, although the performance of the boosting algorithm was somewhat disappointing, its fitness function when used in concert the the genetic algorithm led to results that are worth thinking hard about. 

\section{Conclusions}

Perhaps it helps to restate the challenge. For incidents of mass violence, base rates will be very low. Conventional forms of analysis likely will stumble; our logistic regression results are an instructive example. The choice for research and policy, therefore, is either to abandon serious statistical science or consider rather new approaches to data with low base rates. It this paper, we offer very tentatively a promising option. 

Machine learning (i.e., stochastic gradient boosting) performed far better than the logistic regression, but still fell short of forecasting accuracy that one would ordinarily require (Berk, 2018). When the boosting results were used to construct a fitness function, the genetic algorithm produced outcomes that seem far more interesting. Some predictors, which to our knowledge had never been evaluated before, surfaced as important.  From the hypothetical population of 500, we also learned which of our predictors appeared to perform as aggravators and which appeared to perform as mitigators and how large their impacts were on forecasted risk. 

We used IPV incidents reported to law enforcement as our observational units. Although these data allow for an illustrative application, our results must been seen through the lens of IPV incidents reported to the police. Reported IPV incidents have important similarities to all IPV incidents, but they can be different enough to warrant caution when trying to generalize from reported incidents to all incidents. Whether an IPV incident is reported to authorities introduces a systematic difference, even if potentially small, between the two kinds of incidents. The associations between our predictors and subsequent, violent IPV not reported to law enforcement may differ in strength, and perhaps even in direction.

Although motivated by risk assessment prediction, this paper stops well short of forecasting. In addition, because there is no model of risk, no causal claims can be made. Speculation that some of the associations found may be causal is insufficient. Our intent is to show how a sequence of machine learning algorithms can extract plausible features of a rare population that would not surface from a conventional data analysis alone. In subsequent work, these features might dramatically improve forecasting accuracy for the subset of IPV offenders who pose the greatest risk of injury to victims. In the meantime, they may serve as an instructive checklist of warning signs for violent IPV that at the very least has substantial face validity.

Our methodological conclusions also are offered in a highly provisional manner. As best we can tell, our approach is novel. The challenges inherent in low base rates may require extending existing data analysis tools beyond conventional practice. This paper is a proposal for one way that might be done. 

\section*{References}
\begin{description}
\item
Abramsky, T., Watts, C.H., Garcia-Moreno, C, Devries, K., Kiss, L., Ellsberg, H., Jensen, H., and Heise, L. (2001) ``What Factors Are Associated with Recent Intimate Partner Violence? Findings from the WHO Multi-Country Study on Women's Health and Domestic Violence.'' \textit{BMC Public Health} 11: 109 -- 126.
\item
Berk, R.A. (2007) ``Meta-Analysis and Statistical  Inference'' (with commentary), \textit{Journal of Experimental Criminology} 3(3): 247--297.
\item
Berk, R.A., (2018) \textit{Machine Learning Risk Assessments in Criminal Justice Settings} NewYork: Springer.
\item
Berk, R.~A., Sorenson, S.~B., and He, Y. (2005) Developing a practical forecasting screener for domestic violence incidents. \textit{Evaluation Review} 29(4): 358--382.
\item
Berk, R.A., and Bleich, J.  (2013) ``Statistical Procedures for Forecasting Criminal Behavior: A Comparative Assessment. \textit{Journal of Criminology and Public Policy} 12(3): 515--544, 2013.
\item
Berk, R.A.,  Sorenson, S.B., and Barnes, G. (2016) ``Forecasting Domestic Violence: A Machine Learning Approach to Help Inform Arraignment Decisions.'' \textit{Journal of Empirical Legal Studies} 13(1): 94--115.
Breiman, L. (2001) ``Random Forests.'' {\it Machine Learning} 45: 5--32. 
\item
Burgess, E.~M. (1928) ``Factors Determining Success or Failure on Parole.'' In A.~A. Bruce, A.~J. Harno, E.~.W Burgess, and E.~W., Landesco (eds.) \textit{The Working of the Indeterminate Sentence Law and the Parole System in Illinois} (pp. 205--249). Springfield, Illinois, State Board of Parole. 
\item
Campbell, J.C. \textit{Assessing Dangerousness} Newbury Park: Sage, 1995. 
\item
Campbell, J.C., Webster,D., Koziol-McLain, J., Block, C.,RhD, Doris Campbell, D., Curry. M.A.,. Gary, F. Glass, N., McFarlane, J., Sachs, C., Sharps, P., Ulrich, Y., Wilt, S.A.,  Manganello, J., Xu, X., Schollenberger, J., Frye, V., \& Laughon, K. (2003) Risk factors for femicide in abusive relationships: results from a multisite case control study. \textit{American Journal of Public Health} 93(7): 1089--1097.
\item
Campbell, J. C., Glass, N., Sharps, P. W., Laughon, K., and Bloom, T. (2007) ``Intimate Partner Homicide: Review and Implications of Research and Policy.'' \textit{Trauma, Violence \& Abuse} 8: 246--269.
\item
Campbell, J. C., Webster, D. W., and Glass, N. (2009) ``The Danger Assessment Validation of a Lethality Risk Assessment Instrument for Intimate Partner Femicide.'' \textit{Journal of Interpersonal Violence} 24: 653--674.
\item
Coles, S. (2001) \textit{An Introduction to Statistical Modeling of Extreme Values}. New York: Springer.
\item
Cunha, O. S., and Goncalves, R. A. (2016) ``Predictors of Intimate Partner Homicide in a Sample of Portuguese Male Domestic Offenders.'' \textit{Journal of Interpersonal Violence} doi:10.1177/ 0886260516662304.
\item
Breiman, L. (2001a) ``Random Forests.'' \textit{Machine Learning} 45: 5--32.
\item
Faraway, J.J. (2016) ``Does Data Splitting Improve Prediction?" \textit{Statistics and Computing}, 26 (1-2): 49--60. 
\item
Kaufman, L. and Rousseeuw, P.J. (2005) \textit{Finding Groups in Data: An Introduction to Cluster Analysis.} Wiley, New York.
\item
Hastie, T., Tibshirani, R., and Friedman, J. (2009) \textit{The Elements of Statistical Learning.} Second Edition. New York: Springer.
\item
Luke, S. (2103) \textit{Essentials of Metaheuristics}, second edition, Lulu, available at http://cs.gmu.edu/~sean/book/metaheuristics/.
\item
Mease, D., Wyner, A.J., and Buja, A. (2007) ``Boosted Classification Trees and Class Probability/Quantile Estimation.'' \textit{Journal of Machine Learning Research} 8: 409--439.
\item
Mitchell, M. (1998) \textit{An Introduction to Genetic Algorithms}. Cambridge, MA: MIT Press.
\item
Small, D.S., Sorenson, S.B., and Berk, R.A. (2018) ``After the Gun: Examining Police Visits and Intimate Partner Violence Following Incidents involving a Firearm.'' (under review).
\item
Scrucca, L. (2013) ``GA: A Package for Genetic Algorithms in R.'' \textit{Journal of Statistical Software} 53: 4, http//www.jstatsoft.org/
\item
Spencer, C.M. and Stith, S.M. (2018) ``Risk Factors for Male Perpetration and Female Victimization in Intimate Partner Homicide" A Meta-Analysis.'' \textit{Trauma, Violence, \& Abuse} DOI: 10.1177/1524838018781101.
\item
Straus, M.A. and Gelles, R J. (1990) \textit{Physical Violence in American Families: Risk Factors and Adaptations to Violence in 8,145 Families} New Brunswick, NJ: Transaction.
\item
Storey, J., and Hart, D.D. (2014) ``An Examination of the Danger Assessment as a Victim-Based Risk Assessment Instrument for Lethal Intimate Partner Violence.'' \textit{Journal of Threat Assessment and Management} 1(1): 56 -- 66. 
\item
Umbarkar, A.J., and Sheth, P.D. (2015) ``Crossover Operators in Genetic Algorithms: A Review.'' \textit{ICTACT Journal of Soft Computer} 6(1): 1083--1092 
\item
Weitzman, A. (2018) ``Does Increasing Women's Education Reduce Their Risk of Intimate Partner Violence? Evidence from an Educational Policy Reform.'' \textit{Criminology} 56(3): 575--607.
\end{description}
\end{document}